\documentclass[preprint,onecolumn,showkeys,amssymb,amsmath,aps]{revtex4-1}
\usepackage{graphics}
\usepackage{bm}
\usepackage{amsmath}
\usepackage{amssymb}
\begin{document}

\title{Quantum discord and related measures of quantum correlations in $XY$ chains}
\author{N. Canosa, L. Ciliberti, R. Rossignoli}
\affiliation{Departamento de F\'{\i}sica-IFLP,
Universidad Nacional de La Plata, C.C. 67, La Plata (1900), Argentina}

\begin{abstract}
We examine the quantum correlations of spin pairs in the ground state of finite
$XY$ chains in a transverse field, by evaluating the quantum discord as well as
other related entropic measures of quantum correlations. A brief review of the
latter, based on generalized entropic forms, is also included. It is shown that
parity effects are of crucial importance for describing the behavior of these
measures below the critical field. It is also shown that these measures reach
full range in the immediate vicinity of the factorizing field, where they
become independent of separation and coupling range. Analytical and numerical
results for the quantum discord, the geometric discord and other measures in
spin chains with nearest neighbor coupling and in fully connected spin arrays
are also provided.
\end{abstract}

\maketitle
\section{Introduction}
The last decades have witnessed the great progress experienced by the
interdisciplinary field of quantum information science \cite{NC.00,Ve.06,HR.07},
which began with the recognition of the potential of quantum systems and
quantum correlations for information processing tasks. While it is well known
that quantum entanglement is essential for quantum teleportation \cite{QT.93},
superdense coding \cite{SD.92} and also for achieving exponential speed-up in
pure state based quantum algorithms \cite{JL.03}, the mixed state based quantum
algorithm of Knill and Laflamme \cite{KL.98} showed that such speedup could in
principle be achieved in this case without a substantial presence of
entanglement \cite{DFC.05}. This has oriented the attention to alternative
definitions and measures of quantum correlations for mixed states, like the
quantum discord \cite{Zu.01,HV.01}. While coinciding with the entanglement
entropy for pure states, the quantum discord differs essentially from the
entanglement of formation in the case of mixed states, being non-zero in most
separable mixed states and vanishing just for states which are strictly
classically correlated at least with respect to one of the constituents, i.e.,
diagonal in a standard or conditional product
basis \cite{Zu.01}. The result of Ref.\  \cite{Ca.08} showing the existence of
a finite discord between the control qubit and the remaining qubits in the
circuit of Ref.\  \cite{KL.98}, unleashed a great interest on this measure
and several investigations on its fundamental
properties \cite{Dat.09,SL.09,FA.10,FCOC.11}, on its evaluation on spin chains
and specific states \cite{Dd.08,SA.09,MG.10,WR.10,CRC.10,Luo.08,AR.10,AD.11} as
well as on related measures \cite{Lu.08,WPM.09,Mo.10,DVB.10,RCC.10,RCC.11}, have
been recently made (see Ref.\  \cite{Mo.11} for a recent review). Distinct
quantum capabilities of states with non-zero discord have also been recently
investigated \cite{SKB.11,MD.11,CAB.11,PGA.11}.

Our aim here is to describe the remarkable behavior of the quantum discord and
of other related entropic measures of quantum correlations, in the exact ground
state of finite $XY$ chains in a transverse field \cite{CRC.10}. We first
provide in sections 2--5 a brief review of the quantum discord and of the
generalized entropic measures of quantum correlations discussed in Refs.\
 \cite{RCC.10,RCC.11}. The latter  comprise as particular cases the one-way
information deficit \cite{HH.05,SKB.11} and the geometric measure of discord of
ref.\  \cite{DVB.10}, embedding them in a unified formalism based on
majorization \cite{Wh.78,Bha.97} and general entropic forms \cite{CR.02}. While
their basic features are similar to those of the quantum discord, the
possibility of using simple entropic forms permits an easier evaluation,
allowing for analytical expressions in some cases, as occurs with the geometric
discord of general two qubit states \cite{DVB.10}.

We then use these measures to investigate, in sections 6--8, the quantum
correlations of spin pairs in the exact ground state of finite $XY$ chains in a
transverse field. We review the main results of Ref.\  \cite{CRC.10} on the
behavior of the quantum discord in these chains and also add new results
concerning the behavior of the geometric discord and other related measures 
in such chains. The exact ground 
state of a finite $XY$ chain in a transverse field has a definite spin parity
and this fact will be seen to deeply affect the discord and the previous measures
for fields lower than the critical field $B_c$. We will show that the essential
results in this sector can be interpreted in terms of the discord of mixtures
of aligned pairs.

Moreover, these chains can exhibit a {\it factorizing field}
$B_s$ \cite{Kur.82,Ros.04,Ros.05,Am.06,RCM.08,GAI.09,RCM.09,GAI.10,TT.11}, where
they have a completely separable ground state. For transverse fields, such
eigenstate actually breaks the previous parity symmetry and is hence
degenerate, coinciding $B_s$ in a finite chain with the last crossing of the
two lowest opposite parity levels \cite{RCM.08}. A related remarkable effect is
that in the immediate vicinity of $B_s$, pairwise entanglement, though weak,
reaches {\it full range} \cite{Am.06,RCM.08}, regardless of the coupling range.
Here we will show that the quantum discord as well as the entropic measures of
quantum correlations also reach full range at this point, exhibiting universal
features such as being independent of separation and coupling
range \cite{CRC.10}. Moreover, the value reached by them at this point is
non-negligible and does not decrease with size, in contrast with the pairwise
entanglement, since these measures are not restricted by the monogamy
property \cite{CKW.00} which affects the latter (limiting the
concurrence \cite{Wo.98} to order $n^{-1}$ in an $n$ spin chain if all pairs are
equally entangled). Consequently, the behavior of these measures with the
applied field and separation will deviate significantly from that of the
concurrence or entanglement of formation for $|B|<B_s$. Conclusions are finally
discussed in section 9.

\section{Quantum discord}
The quantum discord was originally defined \cite{Zu.01,HV.01} as the difference
between two distinct quantum versions of the mutual information, or
equivalently, the conditional entropy. For a classical bipartite system $A+B$
described by a joint probability distribution $p_{ij}=p(A=i,B=j)$, the
conditional entropy is defined as the average lack of information about $A$
when the value of $B$ is known: $S(A|B)=\sum_{j}p^B_j S(A|B=j)$, where
$p^B_j=\sum_{i}p_{ij}$ is the probability of outcome $j$ in $B$ and
$S(A|B=j)=-\sum_{i}p_{i/j}\log p_{i/j}$ is the Shannon entropy of the
conditional distribution $p_{i/j}=p_{ij}/p_j^B$. It is a non-negative quantity,
and can also be expressed in terms of the joint entropy
$S(A,B)=-\sum_{i,j}p_{ij}\,\log p_{ij}$ and the marginal entropy $S(B)=-\sum_j
p_j^B\log p_j^B$ as $S(A|B)=S(A,B)-S(B)$. Positivity of $S(A|B)$ then implies
$S(A,B)\geq S(B)$ (and hence $S(A,B)\geq S(A)$) for any classical system.

The last expression for $S(A|B)$ allows a direct quantum generalization, namely
\begin{equation}S(A|B)=S(\rho_{AB})-S(\rho_{B})\,,\label{SC1}\end{equation}
where  $S(\rho)=-{\rm Tr}\,\rho\log_2\rho$ is now the von Neumann entropy and
$\rho_{AB}$ the system density matrix, with $\rho_{B}={\rm Tr}_{A}\,\rho_{AB}$
the reduced state of subsystem $B$. It is well known, however, that Eq.\
(\ref{SC1}) can be negative \cite{Wh.78}, being for instance negative in any
entangled pure state: If $\rho_{AB}^2=\rho_{AB}$, $S(\rho_{AB})=0$ and
$S(A|B)=-E(A,B)$,  where $E(A,B)=S(A)=S(B)$ is the entanglement
entropy \cite{Sch.95}. The positivity of Eq.\ (\ref{SC1}) provides in fact a
basic separability criterion for general mixed states \cite{HHH.96}: $\rho_{AB}$
separable $\Rightarrow$ $S(A|B)\geq 0$, a criterion which can actually be
extended to more general entropies \cite{NK.01,RC.02,RC.03}. We recall that
$\rho_{AB}$ is separable if it can be written as a convex combination of
product states, i.e. $\rho_{AB}=\sum_\alpha q_\alpha
\rho_A^\alpha\otimes\rho_B^\alpha$, with $q_\alpha\geq 0$, $\sum_\alpha
q_\alpha=1$ \cite{RW.89}. 

A second quantum version of the conditional entropy, closer in spirit to the
first classical expression, can be defined \cite{Zu.01} on the basis of a
complete local projective measurement $M_B$ on system $B$ (von Neumann
measurement), determined by one dimensional orthogonal local projectors
$\Pi_j=|j_B\rangle\langle j_B|$. The conditional entropy after such measurement
is
\begin{equation}
S_{M_B}(A|B)=\sum_j p^B_j S(\rho_{A/j})=S(\rho'_{AB})-S(\rho'_B)\,,
 \label{SC2}\end{equation}
where $p^B_j={\rm Tr}\,\rho_{AB} \Pi^B_j$, with $\Pi^B_j=I_A\otimes\Pi_j$, is
the probability of outcome $j$,  $\rho_{A/j}={\rm Tr}_B\,
(\rho_{AB}\Pi_j^B)/p_j^B$ is the reduced state of $A$ after such outcome and
\begin{equation}
\rho'_{AB}=\sum_j p_j^B\rho_{A/j}\otimes \Pi_j=\sum_j \Pi_j^B\rho_{AB}\Pi_j^B\,,
\label{rhp}
\end{equation}
is the average joint state after such measurement, with $\rho'_B={\rm
Tr}_A\,\rho'_{AB}=\sum_j p_j^B\Pi_j$. Eq.\ (\ref{SC2}) represents the average
lack of information about $A$ after the measurement $M_B$ in $B$ is performed
and is clearly non-negative, in contrast with Eq.\ (\ref{SC1}). For a classical
system, both quantities (\ref{SC1})--(\ref{SC2}) are, however, equivalent. We
also mention that for general local measurements, defined by a set of positive
operators $E_j$ ($\sum_j E_j=I$), the first expression in (\ref{SC2}) is to be
used for $S_{M_B}(A|B)$ (with $\Pi_j\rightarrow E_j$).

The quantum discord \cite{Zu.01,HV.01,Ca.08} can then be defined as the
minimum difference between Eqs.
 (\ref{SC2}) and (\ref{SC1}):
  \begin{eqnarray}
D^B(\rho_{AB})
&=&\mathop{\rm Min}_{M_B}[S_{M_B}(A|B)]-S(A|B)
\,,\label{D}
\end{eqnarray}
where the minimization is over all local measurements $M_B$. Due to the
concavity of the von Neumann conditional entropy (\ref{SC1}) with respect to
$\rho_{AB}$, \cite{Wh.78}  Eq.\ (\ref{D}) is non-negative \cite{Zu.01}, vanishing
just for classically correlated states with respect to $B$, i.e. states which
are already of the general form (\ref{rhp}) (a particular case of separable
state) and which remain then unchanged under a specific unread local
measurement. Such states are diagonal in a ``conditional'' product basis
$\{|i_j j\rangle\equiv|{i_{j}}_A\rangle\otimes |j_B\rangle\}$, with
$|{i_{j}}_A\rangle$ the eigenstates of $\rho_{A/j}$.

Eq.\ (\ref{D}) is then non-zero not only in entangled states but also in
separable states not of the form (\ref{rhp}), i.e. those which involve convex
mixtures of non-commuting product states, for which the entanglement of
formation \cite{BDSW.96} vanishes. It is then a measure of all quantum-like
correlations between $A$ and $B$. The distinction with entanglement arises
nonetheless just for mixed states: For pure states $\rho^2_{AB}=\rho_{AB}$,
$S(\rho_{AB})=0$ and $S(\rho'_{AB})=S(\rho'_B)$ for any von Neumann
measurement, reducing the quantum discord exactly to the  entanglement entropy:
$D^A=D^B=E(A,B)$.

Eq.\ (\ref{D}) can of course be also understood \cite{HV.01} as the minimum
difference between the {\it quantum mutual information} \cite{Wh.78}
$I(A:B)=S(A)-S(A|B)=S(\rho_A)+S(\rho_B)-S(\rho_{AB})$, which measures all
correlations between $A$ and $B$ ($I(A:B)\geq 0$, with $I(A:B)=0$ if and only
if $\rho_{AB}=\rho_A\otimes \rho_B$) and the ``classical'' mutual information
$I_{M_B}(A:B)=S(A)-S_{M_B}(A|B)$,
which measures the correlations after the local measurement $M_B$.

\section{Generalized entropic measures of quantum correlations}
Let us now discuss an alternative approach for measuring quantum
correlations \cite{RCC.10}, which allows the direct use of more general entropic
forms. We consider a complete local projective measurement $M_B$ (Von Neumann
type measurement) on part $B$ of a bipartite system initially in a state
$\rho_{AB}$, such that the post-measurement state is given by Eq.\ (\ref{rhp})
if the result is unread. A fundamental property satisfied by the state
(\ref{rhp}) is the majorization relation \cite{Wh.78,Bha.97,NC.00}
\begin{equation}\rho'_{AB}\prec \rho_{AB}\label{m1}\,,\end{equation}
where $\rho'\prec\rho$ means, for normalized mixed states $\rho$, $\rho'$ of
the same dimension $n$,
\[\sum_{j=1}^i p'_j\leq \sum_{j=1}^i p_j,\;\;i=1,\ldots,n-1\,.\]
Here $p'_j$, $p_j$ denote, respectively, the eigenvalues of $\rho'$ and $\rho$
sorted in {\it decreasing} order ($p_j\geq 0$, $\sum_j p_j=1$).  Eq.\
(\ref{m1}) implies that $\rho'_{AB}$ is always {\it more mixed} than
$\rho_{AB}$: If Eq.\ (\ref{m1}) holds, $\rho'_{AB}$ can be written as a convex
combination of unitaries of $\rho_{AB}$: $\rho'_{AB}= \sum_\alpha q_\alpha
U_\alpha\rho_{AB}U_\alpha^\dagger$, with $q_\alpha>0$, $\sum_\alpha q_\alpha=1$
and $U_\alpha^\dagger U_\alpha=I$ \cite{Wh.78,Bha.97,NC.00}.

Eq.\ (\ref{m1}) not only implies that $S(\rho'_{AB})\geq S(\rho_{AB})$ for the
von Neumann entropy, but also
\begin{equation}S_f(\rho'_{AB})\geq S_f(\rho_{AB})\label{m2}\end{equation}
for {\it any}  entropy of the form \cite{CR.02}
\begin{equation}S_f(\rho)={\rm Tr}\,f(\rho)\,,\label{Sf}\end{equation}
where $f:[0,1]\rightarrow\Re$ is a smooth strictly concave function satisfying
$f(0)=f(1)=0$. As in the von Neumann case, recovered for
$f(\rho)=-\rho\log\rho$, these entropies also satisfy $S_f(\rho)\geq 0$, with
$S_f(\rho)=0$ if and only if $\rho$ is a pure state ($\rho^2=\rho$), and
$S_f(\rho)$ maximum for the maximally mixed state $\rho=I_n/n$. Hence, Eq.\
(\ref{m1}) implies a strict disorder increase by measurement which cannot be
fully captured by considering just a single choice of entropy ($S(\rho')\geq
S(\rho)$ does not imply $\rho'\prec\rho$). More generally, Eq.\ (\ref{m1})
actually implies $F(\rho')\geq F(\rho)$ for any Schur concave function $F$ of
$\rho$\cite{Bha.97}. Nonetheless, entropies of the form (\ref{Sf}) are {\it
sufficient} to characterize Eq.\ (\ref{m1}), in the sense that if Eq.\
(\ref{m2}) holds for {\it all} such $S_f$, then $\rho'\prec\rho$ \cite{RC.03}.

We may now consider the generalized information loss due to such
measurement \cite{RCC.10},
\begin{equation}
I_f^{M_B}=S_f(\rho'_{AB})-S_f(\rho_{AB})\,,\label{IfM}
 \end{equation}
which is always {\it non-negative} due to Eqs.\ (\ref{m1})--(\ref{m2}),
vanishing only if $\rho'_{AB}=\rho_{AB}$ due to the strict concavity of $f$.
Eq.\ (\ref{IfM}) is a measure of the
information contained in the off-diagonal elements $\langle i
j|\rho_{AB}|i'j'\rangle$ ($j\neq j'$) of the original state, lost in the
measurement. The minimum of $I_f^{\rm M_B}$ among all complete local
measurements \cite{RCC.10},
\begin{equation}
I_f^B(\rho_{AB})=\mathop{\rm Min}_{M_B} S_f(\rho'_{AB})-S_f(\rho_{AB})
\label{If}\,,\end{equation}
provides then a measure of the quantum correlations between $A$ and $B$ present
in the original state and destroyed by the local measurement in $B$: $I_f^B\geq
0$, vanishing, as the quantum discord (\ref{D}),  only if $\rho_{AB}$ is
already of the form (\ref{rhp}),  i.e., only if it is diagonal in a standard or
conditional product basis.

Again, in the case of a pure state
$\rho_{AB}=|\Psi_{AB}\rangle\langle\Psi_{AB}|$, it can be shown \cite{RCC.10}
that Eq.\ (\ref{If}) reduces to the {\it generalized entanglement entropy}:
$I_f^A=I_f^B=E_f(A,B)$ if $\rho^2_{AB}=\rho_{AB}$, where
$E_f(A,B)=S_f(\rho_A)=S_f(\rho_B)$. The minimizing measurement in this case is
the local Schmidt basis for $|\Psi_{AB}\rangle$:  $M_B=\{|k^B\rangle\langle
k^B|\}$ if $|\Psi_{AB}\rangle=\sum_k \sqrt{p_k}|k^A\rangle\otimes
|k^B\rangle$ \cite{RCC.10}.

In the case of the von Neumann entropy ($S_f(\rho)=S(\rho)$), Eq.\ (\ref{If})
becomes the one-way information deficit \cite{HH.05,SKB.11}, which coincides
with the different version of discord given in the last entry of Ref.\
 \cite{Zu.01} (and denoted as thermal discord in  \cite{Mo.11}). It can be
rewritten in this case in terms of the relative entropy \cite{Wh.78,Ve.02}
$S(\rho||\rho')=-{\rm Tr}\rho(\log\rho' -\log\rho)$ (a non-negative quantity)
as
\begin{equation}
I^{B}(\rho_{AB})\equiv \mathop{\rm Min}_{M_B}S(\rho'_{AB})-S(\rho_{AB})
=\mathop{\rm Min}_{M_B} S(\rho_{AB}||\rho'_{AB})\,,
 \label{I1}\end{equation}
where we have used the fact that the diagonal elements of $\rho_{AB}$ and
$\rho'_{AB}$ in the basis where the latter is diagonal are obviously
coincident. We also note that for these measurements, the quantum discord
(\ref{D})  can be expressed as $D^{B}=\mathop{\rm
Min}_{M_B}[I^{M_B}(\rho_{AB})-I^{M_B}(\rho_{B})]$, coinciding then with $I^{B}$
when the minimizing measurements in (\ref{I1}) and (\ref{D}) are the same and
such that $\rho'_B=\rho_B$.

In the case of the {\it linear entropy} $S_2(\rho)=1-{\rm Tr}\rho^2$, obtained
for $f(\rho)=\rho(1-\rho)$ (i.e., using the linear approximation
$\log\rho\rightarrow\rho-I$ in $-\rho\log\rho$),  Eq.\ (\ref{If})
becomes \cite{RCC.10}
\begin{equation}
I_2^B(\rho_{AB})\equiv \mathop{\rm Min}_{M_B}\,{\rm Tr}(\rho_{AB}^2-{\rho'}_{AB}^2)=
\mathop{\rm Min}_{M_B} ||\rho'_{AB}-\rho_{AB}||^2\,,\label{I2}
\end{equation}
where $||O||^2={\rm Tr}\,O^\dagger O$ is the squared Hilbert Schmidt norm. This
quantity becomes then equivalent to the geometric measure of discord introduced
in Ref.\  \cite{DVB.10}. The latter is defined as the last expression in Eq.\
(\ref{I2}) with minimization over all states diagonal in a product basis, but
the minimum corresponds to a state of the form (\ref{rhp}) \cite{RCC.10}. For
pure states, $I_2^B$ becomes proportional to the squared {\it concurrence}
$C^2_{AB}$, \cite{Wo.98} as for pure states $C^2_{AB}$ is proportional to the
linear entropy of any of the subsystems \cite{Ca.03}.

Finally, in the case of the Tsallis entropy \cite{Ts.09} $S_q(\rho)=(1-{\rm
Tr}\,\rho^q)/(q-1)$, $q>0$, which corresponds to $f(\rho)=(\rho-\rho^q)/(q-1)$,
Eq.\ (\ref{If}) becomes \cite{RCC.11}
\begin{equation} I_q^B(\rho_{AB})=\mathop{\rm Min}_{M_B}
S_q(\rho'_{AB})-S_q(\rho_{AB})\propto
\mathop{\rm Min}_{M_B}{\rm Tr}\,(\rho_{AB}^q-{\rho'}_{AB}^q)\,,
\label{Iq}\end{equation}
with $I_q$ reducing to the one way information deficit for $q\rightarrow 1$ (as
$S_q(\rho)\rightarrow S(\rho)$), and to the geometric discord for $q=2$. This
entropy allows then a simple continuous shift between different measures.

When considering qubit systems, we will normalize entropies such that
$S_f(\rho)=1$ for a maximally mixed two-qubit state $\rho$ (i.e., $2f(1/2)=1$),
implying that all $I_f^B$ will take the value $1$ in a maximally entangled
two-qubit state (Bell state). This implies setting $\log\equiv \log_2$ in the
von Neumann entropy, $S_2(\rho)=2(1-{\rm Tr}\,\rho^2)$ in the linear case (such
that $I_2^B=2{\rm Min}_{M_B}||\rho'_{AB}-\rho_{AB}||^2$) and $S_q(\rho)=(1-{\rm
Tr}\,\rho^{q})/(1-2^{1-q})$ in the Tsallis case.

\section{General stationary conditions for the least disturbing measurement}

The stationary condition $\delta I_f^{M_B}(\rho_{AB})=0$ for the quantity
(\ref{IfM}), obtained by considering a general variation $\delta
|j_B\rangle=(e^{i\delta h_B}-I)|j_B\rangle\approx i\delta h|j_B\rangle$ of the
local measurement basis, where $h_B$ is an hermitian local operator,
reads \cite{RCC.11}
\begin{equation} {\rm Tr}_{A}[f'(\rho'_{AB}),\rho_{AB}]=0\,,
 \label{stif}\end{equation}
i.e., $\sum_{i}[f'(p^i_j)\langle i_j j|\rho_{AB}|i_jk\rangle-f'(p^i_k)\langle
i_k j|\rho_{AB}|i_k k\rangle]=0$, where  $f'$ denotes the derivative of $f$ and
$\langle i_j j|\rho_{AB}|i'_j j\rangle=\delta_{ii'}p^i_j$. In the case of the
geometric discord, $f'(\rho)\propto I-2\rho$ and Eq.\ (\ref{stif}) reduces to
${\rm Tr}_A\,[\rho'_{AB},\rho_{AB}]=0$. In the case of  the quantum  discord
(\ref{D}), Eq.\ (\ref{stif}) should be replaced for these measurements
by \cite{RCC.11}
\begin{equation}
 {\rm Tr}_{A}[f'(\rho'_{AB}),\rho_{AB}]-[f'(\rho'_B),\rho_B]=0\,,
\label{stif2}\end{equation}
with $f(\rho)=-\rho\log\rho$, due to the extra local term.

Eqs.\ (\ref{stif})--(\ref{stif2}) allow us to identify the stationary
measurements, from which the one  providing the absolute minimum of $I_f^{M_B}$
(least disturbing measurement) is to be selected. For instance, if there is a
standard product basis where $\langle
ij|\rho_{AB}|ij'\rangle=\delta_{jj'}p^i_j$ and $\langle
ij|\rho_{AB}|i'j\rangle=\delta_{ii'}p^i_j$, such that the only off-diagonal
elements are $\langle ij|\rho_{AB}|i'j'\rangle$ with $i\neq i'$ and $j\neq j'$,
a measurement in the basis $\{|j_B\rangle\}$ is clearly stationary for {\it
all} $I_f^B$, as Eq.\ (\ref{stif}) is trivially satisfied, leading to a {\it
universal stationary point} \cite{RCC.11}. It will also be stationary for the
quantum discord.

An example of such basis is the Schmidt basis for a pure state,
$|\Psi_{AB}\rangle= \sum_{k=1}^{n_s}\sqrt{p_k}|kk\rangle$, with
$|kk\rangle\equiv|k_A\rangle\otimes |k_B\rangle$ and $n_s$  the Schmidt rank,
since $\langle kl |\rho_{AB}|k'l'\rangle=\delta_{kl}\delta_{k'l'}\sqrt{p_k
p_{k'}}$ for $\rho_{AB}=|\Psi_{AB}\rangle\langle\Psi_{AB}|$. The same holds for
a mixture of a pure state with the maximally mixed state,
\begin{equation}
\rho_{AB}(x)=x|\Psi_{AB}\rangle\langle\Psi_{AB}|+\frac{1-x}{n}I_n\,,
 \label{rhx}\end{equation}
where $n=n_A n_B$ and $x\in[0,1]$. The Schmidt basis provides in fact the {\it
actual minimum} of $I_f^B(x)\equiv I_f^B(\rho_{AB}(x))$ $\forall$ $x\in[0,1]$,
as shown in Ref.\  \cite{RCC.10}. This implies the existence in this case of
a universal least disturbing measurement, and of a concomitant {\it least
mixed} post measurement state, such that $\rho'_{AB}$ majorizes any other
post-measurement state. We can then obtain a closed evaluation of $I_f^B$ for
this case $\forall$ $S_f$, \cite{RCC.10} which shows some of its main features:
\begin{equation}
I_f^B(x)=\sum_{k=1}^{n_s}f(\frac{x(np_k-1)+1}{n})
-f(\frac{x(n-1)+1}{n})-(n_s-1)f(\frac{1-x}{n})
\,.\label{Ifx}
\end{equation}
If $n_s>1$,  it can be shown that $I_f^B(x)>0$ for $x>0$, being a strictly
increasing function of $x$ for $x\in[0,1]$ if $f(p)$ is strictly concave.
Moreover, a series expansion for  small $x$ leads to $I_f^B(x)\approx \alpha
x^2(1-\sum_{k=1}^{n_s}p_k^2)$, where $\alpha=-f''(1/n)/2\geq 0$, indicating a
{\it universal quadratic increase} with increasing $x$ if $f''(1/n)\neq
0$ \cite{RCC.10}. This behavior is then similar to that of the quantum
discord \cite{Zu.01} and quite distinct from that of the entanglement of
formation, which requires a finite threshold value of $x$ for acquiring a
non-zero value.

\section{The two qubit case}
Let us now examine the particular case of a two qubit system. A general
two-qubit state can be written as
\begin{equation}
\rho_{AB}=\frac{1}{4}(I+\bm{r}_A\cdot\bm{\sigma}_A+\bm{r}_B\cdot\bm{\sigma}_B+
 \bm{\sigma}_A^tJ\bm{\sigma}_B)\,,\label{rh2}\end{equation}
where $I\equiv I_2\otimes I_2$ denotes the identity,
$\bm{\sigma}_A=\bm{\sigma}\otimes I_2$ and
$\bm{\sigma}_B=I_2\otimes\bm{\sigma}$. Due to the orthogonality of the Pauli
matrices, we have $\bm{r}_A=\langle \bm{\sigma}_A\rangle$,
$\bm{r}_B=\langle\bm{\sigma}_B\rangle$ and $J=\langle
\bm{\sigma}_A^t\bm{\sigma}_B\rangle$, i.e., $J_{\mu\mu'}=\langle
\sigma_{A\mu}\sigma_{B\mu'}\rangle$, where $\mu,\mu'=x,y,z$ and $\langle
O\rangle={\rm Tr}\,\rho_{AB}\,O$.

A general local projective measurement in this system is just a spin
measurement along a unit vector $\bm{k}$, and is represented by the orthogonal
projectors $\frac{1}{2}(I\pm\bm{k}\cdot\bm{\sigma})$.  Therefore, the most
general post-measurement state (\ref{rhp}) reads
\begin{equation}
\rho'_{AB}=\frac{1}{4}(I+\bm{r}_A\cdot\bm{\sigma}_A+
(\bm{r}_B\cdot\bm{k})\bm{k}\cdot\bm{\sigma}_B+
(\bm{\sigma}_A^tJ\bm{k})(\bm{k}\cdot\bm{\sigma}_B))\,,\label{rhp2}
\end{equation}
and corresponds in matrix notation (setting $\bm{r}$ and $\bm{k}$ as column
vectors) to $\bm{r}_B\rightarrow \bm{k}\bm{k}^t\bm{r}_B$ and $J\rightarrow
J\bm{k}\bm{k}^t$. The general stationary condition (\ref{stif}) can be shown to
lead to the equation \cite{RCC.11} 
 \begin{equation}
\alpha_1\bm{r}_B+\alpha_2J^t\bm{r}_A+\alpha_3J^tJ\bm{k}=\lambda\bm{k}\,,\label{eq}
 \end{equation}
i.e., $\bm{k}\times(\alpha_1\bm{r}_B+\alpha_2J^t\bm{r}_A+\alpha_3J^tJ\bm{k})
=\bm{0}$, which determines the possible values of the minimizing measurement
direction $\bm{k}$. Here
($\alpha_1,\alpha_2,\alpha_3)=\frac{1}{4}\sum\limits_{\nu,\nu'=\pm
1}f'(p^{\nu'}_{\nu})(\nu, \nu\nu'/\lambda_\nu,\nu'/\lambda_\nu)$,  with
$p^{\nu'}_\nu=\frac{1}{4}(1+\nu\bm{r}_B\cdot\bm{k}+\nu'\lambda_{\nu})$ the
eigenvalues of (\ref{rhp2}) and $\lambda_\nu=|\bm{r}_A+\nu J\bm{k}|$. In the
case of the quantum discord (\ref{D}), the additional local term leads to the
modified equation \cite{RCC.11}
 \begin{equation}
 (\alpha_1-\eta)\bm{r}_B+\alpha_2J^t\bm{r}_A+\alpha_3J^tJ\bm{k}=\lambda\bm{k}\,,
 \end{equation}
where here $f(p)=-p\log p$ and $\eta=\frac{1}{2}\sum_{\nu=\pm 1}\nu
f'(p_\nu)=\frac{1}{2}\log(p_-/p_+)$, with $p_\nu=\sum_{\nu'}
p^{\nu'}_\nu=\frac{1}{2}(1+\nu\bm{r}_B\cdot\bm{k})$ the eigenvalues of
$\rho'_B$. A different approach was provided in Ref.\  \cite{AD.11}.

General analytic solutions of these equations can be obtained in a few cases.
For instance, a closed evaluation of $I_f^B$ for any $S_f$ is directly feasible
for any two-qubit state with {\it maximally mixed marginals}, i.e.
\begin{equation}
 \rho_{AB}=\frac{1}{4}(I+\bm{\sigma}_A J\bm{\sigma}_B)\,,\label{Mm}\end{equation}
for which Eq.\ (\ref{eq}) reduces to $J^tJ\bm{k}=\lambda\bm{k}$ $\forall$
$I_f$, indicating that $\bm{k}$ should be an eigenvector of $J^tJ$. Moreover,
it can be shown \cite{RCC.11} that the minimum corresponds to $\bm{k}$ directed
along the eigenvector with the largest eigenvalue of $J^tJ$ $\forall$ $S_f$
(universal least disturbing measurement), such that the post-measurement state
(\ref{rhp})--(\ref{rhp2}) conserves the largest component: By suitable local
rotations, $\bm{\sigma}_A J\bm{\sigma}_B$ can be written as
$\sum_{\mu=x,y,z}J_\mu \sigma_{A\mu}\sigma_{B\mu}$, where $J_\mu$ are the
eigenvalues of $J^tJ$ (the same as those of $JJ^t$), and the least disturbing
measurement leads then to
$\rho'_{AB}=\frac{1}{4}(I+J_{\tilde{\mu}}\sigma_{A\tilde{\mu}}
\sigma_{B\tilde{\mu}}$, where $J_{\tilde{\mu}}={\rm Max}[J_x,J_y,J_z]$. The
final result for $I_f^B$ (obviously identical to $I_f^A$ for this state)
is \cite{RCC.11}
\begin{equation}
I_f^B(\rho_{AB})=2f(\frac{p_1+p_2}{2})+2f(\frac{p_3+p_4}{2})-f(p_1)-f(p_2)
-f(p_3)-f(p_4)\,,
\label{Ifxx}
\end{equation}
where $(p_1,p_2,p_3,p_4)$ are the eigenvalues of $\rho_{AB}$ sorted in
decreasing order ($p_{1,2}=\frac{1+J_z\pm(J_x-J_y)}{4}$, $p_{3,4}=
\frac{1-J_z\pm(J_x+J_y)}{4}$ if $|J_z|\geq |J_x|\geq |J_y|$ and $J_z\geq 0$,
$J_x\geq 0$).  It is verified that $I_f^B=0$ only if $p_1=p_2$ {\it and}
$p_3=p_4$, in which case $\rho_{AB}=\rho'_{AB}$ is a classically correlated
state \cite{RCC.11}. In the von Neumman case, Eq.\ (\ref{Ifxx}) is just the
quantum discord for this state, as in this case it coincides with the one-way
information deficit ($\rho'_B=\rho_B$ are maximally mixed). In the case of the
linear entropy, Eq.\ (\ref{Ifxx}) yields the geometric discord and reduces
to \cite{RCC.11} $I_2^B(\rho_{AB})=(p_1-p_2)^2+(p_3-p_4)^2$.

The linear entropy case (\ref{I2}) is obviously the most simple to evaluate,
and in this sense the most convenient. A full analytic evaluation for a general
two-qubit state was achieved  in Ref.\   \cite{DVB.10}. Since ${\rm
Tr}\sigma_\mu\sigma_{\mu'}=2\delta_{\mu\mu'}$, one easily obtains in this case
\begin{eqnarray}
 I_2^{\bm{k}}&=&S_2(\rho'_{AB})-S_2(\rho_{AB})
 =\frac{1}{2}({\rm tr}\,M_2-\bm{k}^tM_2\bm{k})\,,\;\;M_2=\bm{r}_B\bm{r}_B^t+J^tJ\,,
 \end{eqnarray}
where $||J||^2={\rm tr}J^tJ$, $|\bm{r}|^2=\bm{r}\cdot\bm{r}$ and $M_2$ is a
positive semidefinite symmetric matrix. The minimum of $I_2^{\bm{k}}$ is then
obtained when $\bm{k}$ is directed along the eigenvector associated with the
maximum eigenvalue $\lambda_1$ of $M_2$, leading to \cite{DVB.10}
\begin{equation}I_2^B=\mathop{\rm Min}_{\bm{k}}
I_2^{\bm{k}}=\frac{1}{2}({\rm tr}\,M_2-\lambda_1)\,.\label{I2s}\end{equation}
It is easily seen that Eq.\ (\ref{eq}) reduces in this case to the eigenvalue
equation $M_2\bm{k}=\lambda\bm{k}$, such that the stationary directions are
those of the eigenvectors of $M_2$.

Similarly, the $q=3$ case in the Tsallis entropy, $S_3(\rho)\propto (1-{\rm
Tr}\,\rho^3)$, can also be fully worked out analytically \cite{RCC.11}. We
obtain
\begin{eqnarray}
 I_3^{\bm{k}}&=&S_3(\rho'_{AB})-S_3(\rho_{AB})=\frac{1}{4}({\rm tr}\,M_3
-2{\rm det}\,J-\bm{k}^tM_3\bm{k})\,,\\
M_3&=&\bm{r}_B\bm{r}_B^t+J^tJ+\bm{r}_B\bm{r}_A^tJ+J^t\bm{r}_A\bm{r}_B^t\,,
\end{eqnarray}
where $M_3$ is again a positive semidefinite symmetric matrix, with ${\rm
tr}M_3=|\bm{r}_B|^2+||J||^2+2\bm{r}_A^tJ\bm{r}_B$. Its minimum corresponds then
to $\bm{k}$ along the eigenvector with the maximum eigenvalue $\lambda_1$ of
$M_3$, which leads to \cite{RCC.11}
\begin{equation}I_3^B=\mathop{\rm Min}_{\bm{k}}I_3^{\bm{k}}=\frac{1}{4}({\rm tr}
\,M_3-2{\rm det}\,J-\lambda_1)\,.\label{I3s}\end{equation} It is again verified
that Eq.\ (\ref{eq}) leads here to the same eigenvalue equation
$M_3\bm{k}=\lambda\bm{k}$, as
$(\alpha_1,\alpha_2,\alpha_3)=(\bm{r}_B^t+\bm{r}_A^tJ,\bm{r}_B^t\bm{k},1)$. In
the case of the state (\ref{Mm}), we obtain \cite{RCC.11}
$I_3^B=(p_1-p_2)^2(p_1+p_2)+(p_3-p_4)^2(p_3+p_4)$.

It should be stressed that for the case of two qubits, these two entropies,
$S_2$ and $S_3$, lead to {\it the same entanglement monotone} \cite{Vi.00},
since for an arbitrary single qubit state they become
identical: \cite{RCC.10,RCC.11} $S_2(\rho_A)=S_3(\rho_A)=1-|\bm{r}_A|^2$ for
$\rho_A=\frac{1}{2}(I_2+\bm{r}_A\cdot\bm{\sigma})$. Both quantities $I_2^B$ and
$I_3^B$ reduce then to the standard squared concurrence \cite{Wo.98} $C^2_{AB}$
in the case of a pure two-qubit state.

\section{The case of a mixture of two aligned states}
We are now in a position to examine the important case of a mixture of two
aligned spin $1/2$ states \cite{CRC.10}, which will allow us to understand the
behavior of the quantum discord of spin pairs in finite ferromagnetic $XY$
chains, particularly in the vicinity of the transverse factorizing field
 \cite{Kur.82,Ros.04,Ros.05,Am.06,RCM.08,GAI.09,RCM.09,GAI.10,CRC.10,TT.11}. We
consider the bipartite state
\begin{eqnarray}
\rho_{AB}(\theta)&=&\frac{1}{2}(|\theta\theta\rangle\langle\theta\theta|
+|-\theta-\theta\rangle\langle-\theta-\theta|)
\label{st}\\
&=&\frac{1}{4}(I+\cos\theta(\sigma_{Az}+\sigma_{Bz})
+\cos^2\theta\,\sigma_{Az}\sigma_{Bz}+\sin^2\theta\,\sigma_{Ax}\sigma_{Bx})\,,
\label{st2}
\end{eqnarray}
where $|\theta\rangle=e^{-i\theta\sigma_y/2}|0\rangle$ denotes the state with
its spin aligned along $\bm{k}=(\sin\theta,0,\cos\theta)$, and (\ref{st2})
corresponds to the spin $1/2$ case, where
$|\theta\rangle=\cos\frac{\theta}{2}|0\rangle+\sin\frac{\theta}{2}|1\rangle$
and we have used the notation of Eq.\ (\ref{rh2}). It is a particular case of
$X$ state \cite{AR.10}, i.e., states that commute with the $S_z$ parity
$P_z=-e^{i\pi(\sigma_{Az}+\sigma_{Bz})/2}$ \cite{RCC.11}.

The state (\ref{st}) arises, for instance, as the reduced state of {\it any}
pair in the $n$-qubit pure states
\begin{equation}
|\Theta_{\pm}\rangle=
\frac{|\theta\theta\ldots\theta\rangle\pm|-\theta\ldots-\theta\rangle}
{\sqrt{2(1\pm\langle-\theta|\theta\rangle^{n})}}\,,\label{sts}
\end{equation}
if the complementary overlap $\langle -\theta|\theta\rangle^{n-2}$
($\langle-\theta|\theta\rangle=\cos\theta$ for spin $1/2$) can be neglected
(i.e., $n$ large and $\theta$ not too small). As will be seen in the next
sections, the states (\ref{sts}) are the actual exact ground states of such
chains in the immediate vicinity of the factorizing field.

\begin{figure}
\vspace*{-0.25cm}

\centerline{\scalebox{.65}{\includegraphics{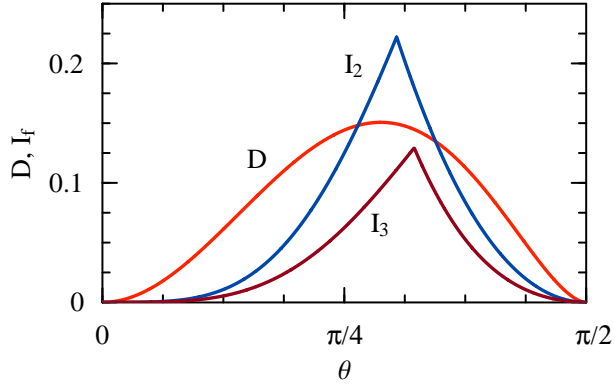}}}
 \vspace*{-0.25cm}
\caption{Quantum correlation measures in the mixture of aligned states
(\ref{st}): The quantum discord $D$, the geometric discord $I_2$ and the
``cubic'' discord $I_3$, as a function of the angle $\theta$. Normalization is
such that all measures take the value $1$ in a maximally entangled two-qubit
state. Due to the symmetry of the state, $D=D^A=D^B$ and $I_f=I_f^A=I_f^B$
$\forall$ $f$.} \label{f1}
\end{figure}

The state (\ref{st}) is clearly separable, i.e., a convex combination of
product states \cite{RW.89}, but is classically correlated, i.e., diagonal in a
product basis, just for $\theta=0$ or $\theta=\pi/2$. Accordingly, both the
quantum discord and all measures (\ref{If}), including the geometric discord,
will be non-zero just for $\theta\in(0,\pi/2)$. As seen in Fig.\ \ref{f1}, they
all exhibit similar  qualitative features, although significant differences
concerning the minimizing measurement arise. Due the symmetry of the state, it
is apparent that $D^A=D^B=D$ and $I_f^A=I_f^B=I_f$ $\forall$ $\theta$.

It is rapidly seen from Eq.\ (\ref{eq}) that for this state (as well as any
other $X$ state), spin measurements along $x$, $y$ or $z$ {\it are stationary},
for both the quantum discord and all measures $I_f$ \cite{RCC.11}. In the case
of the quantum discord, the minimizing measurement for this state (which is of
rank $2$, and hence minimized through a standard von Neumann
measurement \cite{Mo.11}) is in fact along $x$ $\forall$ $\theta\in(0,\pi/2)$,
in which case the eigenvalues of $\rho'_{AB}$ become
$p^{\nu'}_\nu=\frac{1}{4}(1+\nu'\sqrt{\cos^2\theta+\sin^4\theta})$, being
twofold degenerate. The final result for the quantum discord can then be
expressed as \cite{CRC.10}
\begin{equation}
 D=\sum_{\nu=\pm 1} [2f(\frac{1+\nu\sqrt{1-\frac{1}{4}\sin^2 2\theta}}{4})
-f(\frac{1+\nu\cos^2\theta}{2})+f(\frac{1+\nu\cos\theta}{2})]-1\,,\label{Dth}
\end{equation}
where $f(p)=-p\log_2 p$. It is maximum at $\theta\approx 1.15\pi/4$. For
$\theta\approx 0$, $D$ vanishes quadratically ($D\propto\theta^2$) whereas for
$\theta\rightarrow\pi/2$, $D\propto
(\frac{\pi}{2}-\theta)^2(-\log_2(\frac{\pi}{2}-\theta)^2+c)$.

On the other hand, the geometric discord (\ref{I2}) and the ``cubic'' discord
($q=3$ in Eq.\ (\ref{Iq})) can be directly evaluated using Eqs.\
(\ref{I2s})--(\ref{I3s}). We obtain \cite{RCC.11}
\begin{eqnarray}
I_2&=&\left\{\begin{array}{lr}
\frac{1}{2}\sin^4\theta&\;\;\;\theta<\theta_{c2}\\\frac{1}{2} \cos^2\theta+\cos^4\theta
&\;\;\;\theta>\theta_{c2}\end{array},\right. \label{I2th}\\
I_3&=&\left\{\begin{array}{lr}\frac{1}{4}\sin^4\theta&\;\;\;\theta<\theta_{c3}\\
\frac{1}{4}(\cos^2\theta+3\cos^4\theta) &\;\;\;\theta>\theta_{c3}\end{array},\right.
\label{I3th}
 \end{eqnarray}
where $\cos^2\theta_{c2}=1/3$ ($\theta_{c2}\approx 0.61\pi/2$) and
$\cos^2\theta_{c3}=(\sqrt{17}-3)/4$ ($\theta_{c3}\approx 0.64\pi/2$), the
minimizing measurement direction changing abruptly from $z$ to $x$ at
$\theta=\theta_c$ as $\theta$ increases,  in contrast with the quantum discord.
Both $I_2$ and $I_3$ exhibit therefore a cusp like maximum at
$\theta=\theta_c$, as seen in Fig.\ \ref{f1}. It is also seen from Eqs.\
(\ref{I2th})--(\ref{I3th}) that these quantities vanish as $\theta^4$ for
$\theta\rightarrow 0$,  whereas for $\theta\rightarrow\pi/2$ they vanish
quadratically ($\propto (\pi/2-\theta)^2$). For completeness, it should be also
mentioned that the behavior of the least disturbing local measurement for this
state depends actually on the choice of entropy. For instance, in the von
Neumann case (where Eq.\ (\ref{If}) becomes the one-way information deficit
(\ref{I1})), we obtain instead a smoothed $z\rightarrow x$ transition for the
minimizing measurement direction, which evolves continuously from $z$ to $x$ in
a small intermediate interval \cite{RCC.11}.

\section{XYZ spin chains and transverse factorizing field}
Let us now use the previous measures and results to analyze the quantum
correlations between spin pairs in a chain of spins $s_i$. We will consider
finite chains with $XYZ$ couplings of arbitrary range immersed in a transverse
magnetic field, not necessarily uniform, such that the Hamiltonian reads
\begin{equation}
H=\sum_i B^i s_{iz}-\frac{1}{2}\sum_{\mu=x,y,z}\sum_{i,j}
J^{ij}_\mu s_{i\mu}s_{j\mu}\,,
\label{H}
\end{equation}
where $s_{i\mu}$ denote the components of the local spin $\bm{s}_i$ (assumed
dimensionless). We first remark that the Hamiltonian (\ref{H}) always commutes
with the total $S_z$ parity or phase flip
\begin{equation}
P_z=\otimes_{i=1}^n \exp[-i\pi (s_{iz}-s_i)]\,,\label{Pz}
\end{equation}
irrespective of the coupling range, anisotropy, geometry or dimension of the
array. Hence, non-degenerate eigenstates will have a definite parity. In fact,
the ground state of finite chains will typically exhibit a series of parity
transitions as the field increases from $0$, before ending in an almost aligned
state for sufficiently large fields.

A related remarkable effect in these chains is the possibility of exhibiting
{\it a completely separable} exact eigenstate at a {\it factorizing field}. The
existence of a factorizing field was first discussed in Ref.\  \cite{Kur.82},
and its properties together with the general conditions for its existence were
recently analyzed in great detail by several
authors \cite{Ros.04,Ros.05,Am.06,RCM.08,GAI.09,RCM.09,GAI.10,CRC.10,TT.11}. At
the {\it transverse} factorizing field, finite $XYZ$ chains actually exhibit a
{\it pair} of completely separable and degenerate parity breaking exact
eigenstates \cite{RCM.08,RCM.09}, which can be ground states under quite general
conditions. In such a case the transverse factorizing field corresponds  to a
ground state parity transition (typically the last parity
transition \cite{RCM.08,RCM.09}), where the lowest energy levels of each parity
subspace cross and enable the formation of such eigenstates. Let us notice that
while these lowest levels become practically degenerate in a large chain for
fields $|B|<B_c$, they are not exactly degenerate in a finite chain, except at
crossing points \cite{RCM.08,RCM.09,PF.09,CR.07}.

Let us then first describe the general conditions for which a separable parity
breaking state of the form
\begin{equation}
|\Theta\rangle=|\theta_1\ldots\theta_n\rangle=\otimes_{j=1}^n
 \exp[-i\theta_js_{jy}]|0_j\rangle\,,\label{stt}
\end{equation}
where $s_{jz}|0_j\rangle=-s_j|0_j\rangle$, can be an {\it exact} eigenstate of
(\ref{H}). By inserting  (\ref{stt}) in the equation
$H|\Theta\rangle=E|\Theta\rangle$, it can be shown that such conditions
are \cite{RCM.09}
\begin{eqnarray}
J_y^{ij}&=&J_x^{ij}\cos\theta_i\cos\theta_j+J_z^{ij}
\sin\theta_i\sin\theta_j\,,
\label{aa}\\B^i\sin\theta_i&=&\sum_{j}
(s_j-\frac{1}{2}\delta_{ij})(J_x^{ij}\cos\theta_i\sin\theta_j
-J_z^{ij}\sin\theta_i\cos\theta_j) \,,\label{bb}
\end{eqnarray}
which are valid for arbitrary spins $s_i$. They determine, for instance, the
values of $J_y^{ij}$ and $B^i$ in terms of $J_x^{ij}$, $J_z^{ij}$, $s_i$ and
$\theta_i$. A careful engineering of couplings and fields can then always
produce a chain with such eigenstate, for any chosen values of $\theta_i$. It
is also apparent that this eigenstate is degenerate, since
$P_z|\Theta\rangle=|-\Theta\rangle=\otimes_{j=1}^n
\exp[i\theta_js_{jy}]|0_j\rangle$ will have the same energy (and differ from
$|\Theta\rangle$ if $\sin\theta_j\neq 0$ for some $j$),  indicating that these
fields necessarily correspond to the crossing of two opposite parity levels.
Each local state in the product (\ref{stt}) is a local coherent state. We also
note that any state  $\otimes_{j=1}^n
e^{-i\bm{\theta}_j\cdot\bm{s}_j}|0_j\rangle$ can be written, except for a
normalization factor, in the form (\ref{stt}) by allowing a complex
$\theta_j$ \cite{RCM.09}. Eqs.\ (\ref{aa})--(\ref{bb}) are then generally valid
for such type of states.

The second equation (\ref{bb}) cancels the matrix elements of $H$ between
$|\Theta\rangle$ and one spin excitations, and is then a ``mean field-like''
equation, i.e., that which arises when minimizing the average energy $\langle
\Theta|H|\Theta\rangle$ with respect to the $\theta_i$, for fixed fields and
couplings.  The first equation (\ref{aa}) ensures that the minimizing separable
state is an exact eigenstate, by cancelling the residual matrix elements of $H$
connecting $|\Theta\rangle$ with the remaining states (two-spin excitations).
It can be also shown \cite{RCM.09} that in the ferromagnetic-type case
\begin{equation}|J_y^{ij}|\leq J_x^{ij}\;\;\forall\;i,j\,,\end{equation}
where all off-diagonal elements of $H$ in the standard basis of $s_z$
eigenstates are real and negative, the state (\ref{stt}) is necessarily a {\it
ground state} if $\theta_j\in(0,\pi/2)$ $\forall$ $j$, as the  exact ground
state  must have (or can be chosen to have if degenerate) expansion
coefficients of the same sign in this basis (different signs will not decrease
$\langle H\rangle$), and hence cannot be orthogonal to
$|\Theta\rangle$ \cite{RCM.08,RCM.09}.

In particular, a uniform solution $\theta_j=\theta$ $\forall$ $j$, leading to
$|\Theta\rangle=|\theta\ldots\theta\rangle$, is feasible if the coupling
anisotropy
\begin{equation}
\chi=\frac{J_y^{ij}-J_z^{ij}}{J_x^{ij}-J_z^{ij}}\,,\label{chi}
\end{equation}
is {\it constant} $\forall$ $i,j$ and non-negative \cite{RCM.09}.  Of course, if
$\chi>1$ we can change it to $\chi\in(0,1)$ by swapping $x\leftrightarrow y$
through a rotation of $\pi$ round the $z$ axis. In such a case, Eqs.\
(\ref{aa})--(\ref{bb}) lead to
\begin{eqnarray}\cos^2\theta&=&\chi\,,\label{aaa}\\
B^i&=&\sqrt{\chi}\sum_j (J_x^{ij}-J_z^{ij}))(s_j-\frac{1}{2}\delta_{ij})\,,
\label{bbb}\end{eqnarray}
where Eq.\ (\ref{bbb}) holds for $\sin\theta\neq 0$.

Eqs. (\ref{aaa})-(\ref{bbb}) allow, for instance, the existence of a
factorizing field for uniform first neighbor couplings
$J_\mu^{ij}=J_\mu\delta_{j,i\pm 1}$ in a finite linear spin $s$ chain if
$\chi=\frac{J_y-J_z}{J_x-J_z}>0$, both in the cyclic case ($J_\mu^{1n}=J_\mu$),
where the factorizing field is completely {\it uniform},
\begin{equation}B^i=B_s=2s\sqrt{\chi}(J_x-J_z)\,,\label{Bs}\end{equation}
as well as in the open case ($J_{\mu}^{1n}=0$), where $B^i=B_s$ at inner sites
but $B^1=B^n=B_s/2$ at the borders \cite{RCM.09}. A fully and equally connected
spin $s$ array with $J_\mu^{ij}=2J_\mu/(n-1)$ $\forall$ $i\neq j$ (Lipkin
model \cite{VPM.04,DV.05})  will also exhibit a uniform transverse factorizing
field at $B=B_s$ if $\chi>0$ \cite{RCM.08,MRC.08}. The ensuing state
$|\Theta\rangle$ can be ensured to be a {\it ground state} in all these cases
if $|J_y|\leq J_x$ (when $\chi\in[0,1]$).  Other possibilities, like solutions
with alternating angles \cite{RCM.09}, can also be considered.

\section{Quantum correlations in the definite parity ground states}

Let us now focus on finite spin chains which exhibit a separable parity
breaking exact eigenstate $|\Theta\rangle$ at the factorizing field $B_s$.
It will be of course degenerate with $|-\Theta\rangle=P_z|\Theta\rangle$.
The important point is that the definite parity states
\begin{equation}
|\Theta_{\pm}\rangle
=\frac{|\Theta\rangle\pm|-\Theta\rangle}{\sqrt{2(1\pm\langle-
\Theta|\Theta\rangle)}}
\,,\label{PT}
\end{equation}
i.e. the states (\ref{sts}) in the uniform case, will also be exact ground
states at $B_s$. Moreover, since the exact ground state of a finite chain will
actually be non-degenerate away from the factorizing field (and the other
crossing points), it will have a definite parity. Hence, the actual ground
state side-limits at $B=B_s$ will be given by the definite parity states
(\ref{PT}) (rather than  $|\pm\Theta\rangle$). A ground state parity transition
$-$ $\rightarrow$ $+$ will then take place as the field increases across
$B_s$ \cite{RCM.08,RCM.09}.

While the ground states of each parity sector become degenerate in the large
$n$ limit for fields $|B|<B_s$, in finite chains the degeneracy is lifted and
the actual ground state will exhibit important correlations arising just from
the definite parity effect. For instance, in the immediate vicinity of $B_s$,
the pairwise correlations, rather than vanish, will approach the values
determined by the states (\ref{PT}). They will then depend on $\theta_i$ and
$\theta_j$ for a pair $i,j$, irrespective of the separation $i-j$. In the
uniform case, such correlations will then be {\it independent of the
separation}, since the states (\ref{PT}) will be completely symmetric and will
lead to a separation independent pair reduced state $\rho_{ij}={\rm
Tr}_{\bar{ij}}|\Theta_{\pm}\rangle\langle\Theta_{\pm}|$ ($\bar{ij}$ denotes the
rest of the chain). Such state is given exactly by \cite{CRC.10}
\begin{eqnarray}
\rho_{ij}^\varepsilon(\theta)&=&\frac{|\theta\theta\rangle\langle\theta\theta|+
|\!-\!\theta\!-\!\theta\rangle\langle\!-\!\theta\!-\!\theta|+
\varepsilon(|\theta\theta\rangle\langle\!-\!\theta\!-\!\theta|
+|\!-\!\theta\!-\!\theta\rangle\langle\theta\theta|)}
{2(1+\varepsilon\langle\theta\theta|\!-\!\theta\!-\!\theta\rangle)}
 \label{ste}\end{eqnarray}
where $\varepsilon=\pm\cos^{n-2}\theta$ for $s=1/2$. This parameter is then
small for not too small $n$ and $\theta$. In the $\varepsilon\rightarrow 0$
limit one recovers the mixture (\ref{st}).

The concurrence of the state (\ref{ste}) (a measure of its
entanglement \cite{Wo.98}) depends essentially on $\varepsilon$ and is therefore
small, vanishing for $\varepsilon\rightarrow 0$ (as in this limit the state
(\ref{ste}) becomes separable). It is  given  explicitly
by \cite{CRC.10,RCM.08,RCM.09}
\begin{equation}
C={\frac{|\varepsilon|\sin^2\theta}{1+\varepsilon\cos^2\theta}}\,,\label{C2}
\end{equation}
which is parallel (antiparallel) \cite{Am.06} for $\varepsilon>0$ ($<0$). Its
maximum value is  $2/n$ (in agreement with the monogamy property \cite{CKW.00}),
reached for $\theta\rightarrow 0$ in the negative parity case, where the state
(\ref{sts}) approaches an $W$-state \cite{RCM.08}.

In contrast, we have seen that the quantum and geometric discord, as well as
the other measures of quantum correlations (\ref{If}), do acquire finite and
non-negligible values in the mixture (\ref{st}) (Fig.\ \ref{f1}), i.e., in the
state (\ref{ste}) even for $\varepsilon\rightarrow 0$, entailing {\it
simultaneous and coincident} finite values for all pairs in the state
(\ref{sts}) \cite{CRC.10}. This implies in turn {\it infinite range} of pairwise
quantum correlations, as measured by $D$ or $I_f$, at least in the immediate
vicinity of the factorizing field $B_s$, where they will be described by the
state (\ref{st}) and will therefore be {\it independent  of both separation and
coupling range}.

\begin{figure}
\vspace*{-0.25cm}

\centerline{\scalebox{.65}{\includegraphics{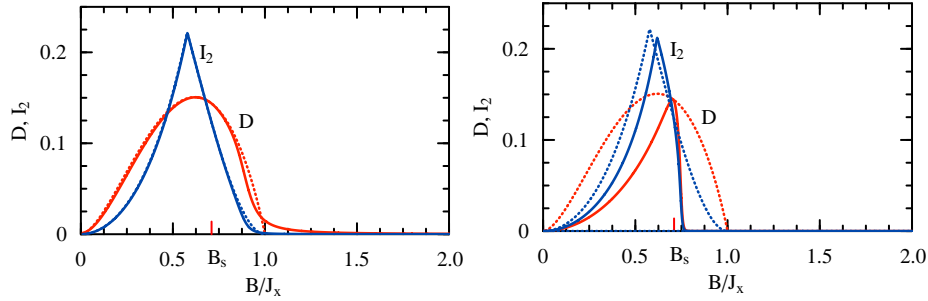}}}
 \vspace*{-0.25cm}

\caption{Left: The quantum discord $D$ and the geometric discord $I_2$ between
spin pairs in the exact ground state of a fully connected $XY$ spin $1/2$ array
of $n=50$ spins  with coupling anisotropy $\chi=J_y/J_x=0.5$, as a function of
the scaled transverse applied field $B$. The dotted lines depict the result for
the mixture of aligned states (\ref{st}) at the mean field angle
$\cos\theta=B/J_x$, which is almost coincident with the exact result for
$|B|<J_x$ and exactly coincident at the factorizing field $B_s=\sqrt{\chi}J_x$.
Right: Same quantities for a cyclic chain of $n=50$ spins with first neighbor
$XY$ couplings with the same anisotropy, for a distant pair ($L=25$). The
dotted lines depict again the results for the mixture of aligned states
(\ref{st}) at the mean field angle, which now coincide with the exact results
only at the factorizing field $B_s$.} \label{f2}
\end{figure}

As illustration, we first depict in Fig.\ \ref{f2} the quantum discord and the
geometric discord of spin $1/2$ pairs for $XY$ couplings ($J_z^{ij}=0$
$\forall$ $i,j$) in the ground state of the fully connected array (Lipkin-type
model) and of the nearest-neighbor cyclic chain, for anisotropy $J_y/J_x=0.5$
in the ferromagnetic type case ($J_x>0$). The emergence of a finite appreciable
value of these quantities for $|B|<B_c$, persisting for pairs with large
separation even in the case of first neighbor couplings, is then a direct
consequence of the definite parity effect. The exact results for $n=50$ spins
were computed by direct diagonalization of $H$ in the Lipkin case (where the
ground state belongs to the completely symmetric representation having total
spin $S=n/2$), while in the nearest neighbor  chain they were obtained through
the exact Jordan-Wigner fermionization \cite{LM.61}, taking into account the
parity effect exactly in the discrete Fourier transform  (see for instance
Refs.\  \cite{CR.07,RCM.08,PF.09}). In both cases the ground state exhibits
$n/2$ parity transitions as the field increases from $0$, the last one at
$B_s$, although for the case depicted ($N=50$), their effects on $D$ or $I_2$
are not visible in the scale  of the figure (they become visible for smaller
$n$  \cite{CRC.10}).

It is verified that at the factorizing field (\ref{Bs}), the exact results for
$D$ and $I_2$ in both models coincide with those obtained for the mixture
(\ref{st}), i.e., with Eqs.\ (\ref{Dth}) and (\ref{I2th}) for
$\cos\theta=\sqrt{\chi}$ (Eq.\ (\ref{aaa})), being then identical and the same
for any pair at this point.  Moreover, in the fully connected case the exact
results for $D$ and $I_2$ are actually practically coincident with those
obtained from the mixture (\ref{st}) (dotted lines) {\it in the whole region
$|B|<J_x$} if $\theta$ is the mean field angle satisfying $\cos\theta=B/J_x$
(Eq.\ (\ref{bb})), since the ground state is in this region well approximated
by the definite parity states (\ref{PT}) or (\ref{sts}), even away from $B_s$.
The behavior of $D$ and $I_2$ for $B\in[0,J_x]$ resembles then that obtained
for the mixture (\ref{st}) for $\theta\in[0,\pi/2]$. This is not the case in
the chain with first neighbor couplings, where the agreement holds just at
$B_s$, and where $D$ and $I_2$ become appreciable only for
$|B|<B_c=(J_x+J_y)/2$ (which is smaller than the mean field critical field
$J_x$ but slightly above the factorizing field $\sqrt{\chi}$). Nonetheless, the
values attained by $D$ and $I_2$ in the whole region $|B|<B_s$ are still quite
large, owing to the definite parity effect, although they exhibit the effects
of correlations beyond the parity projected mean field description provided by
the states (\ref{PT}).

\begin{figure}[t]
\vspace*{-0.25cm}

\centerline{\scalebox{.65}{\includegraphics{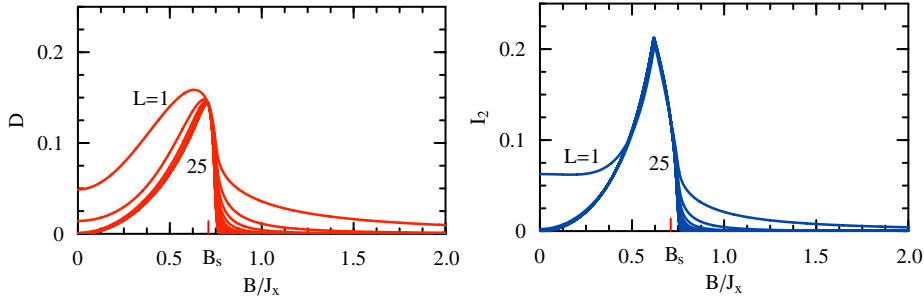}}}
 \vspace*{-0.25cm}
 
\caption{The quantum discord $D$ (left) and the geometric discord $I_2$ (right)
between spin pairs with separation $L=1,2,\ldots,n/2$ in the exact ground state
of a cyclic chain of $n=50$ spins with first neighbor $XY$ couplings and
anisotropy $\chi=0.5$. The results for all separations are simultaneously
depicted. They all merge at the factorizing field $B_s$, where they coincide
with the result for the mixture (\ref{st}) with
$\cos\theta=B_s/J_x=\sqrt{\chi}$.} \label{f3}
\end{figure}

We also depict in Fig.\ \ref{f3} results for $D$ and $I_2$ for all separations
$L=1,\ldots,n/2$ of the spins of the pair, in the nearest neighbor coupling
case (in the fully connected model they are obviously identical $\forall$ $L$).
It is seen that the values of $D$ and $I_2$ (and the same for $I_3$ or other
$I_f$'s) rapidly saturate as $L$ increases in the region $|B|<B_c$, reaching
here a finite non-negligible value due to the definite parity effect, whereas
for $|B|>B_c$ they are appreciable just for the first few neighbors ($L=1,2$).
In the last region they can be  described perturbatively \cite{CRC.10}.

It is apparent from Figs.\ \ref{f2}--\ref{f3} that the same qualitative
information can be obtained either from the quantum discord $D$ or the
geometric discord $I_2$, except for the type of maximum. That of $I_2$ is
cusp-like due to the sharp $x\rightarrow z$ transition in the minimizing
measurement direction that arises as the field increases, which parallels that
occurring for the state (\ref{st}) as $\theta$ decreases (see Fig.\ \ref{f1}).
Such transition reflects the change in the type of pairwise correlation, and
resembles that of the concurrence (which changes from antiparallel to parallel
at $B_s$ \cite{RCM.08}). We also mention that while the behavior of $I_3$ (not
shown) is similar to $I_2$, other $I_f$ can exhibit a smoothed maximum as the
transition from $x$ to $\rightarrow z$ in the measurement direction can be
continuous \cite{RCC.11}.

We finally remark that the exact ground state pairwise concurrence in the fully
connected case is small (of order $n^{-1}$ and bounded above by
$2/n$), \cite{MRC.08} such that the entanglement monotones associated with $D$
and $I_2$ (the entanglement of formation and the squared concurrence) are very
small in the scale of fig.\ \ref{f2}. The same occurs with the concurrence of
largely separated pairs in the nearest neighbor case \cite{ON.02}, which is
non-zero (but very small for this anisotropy and size) just in the immediate
vicinity of $B_s$ \cite{RCM.08,RCM.09}.

\section{Conclusions}
We have examined the behavior of pairwise quantum correlations in the exact
ground state of finite ferromagnetic-type $XY$ spin chains in a transverse
field, by analyzing the quantum discord as well as other generalized measures of
quantum correlations. We have first provided a brief review of the latter,
which are based on general entropic forms and defined as the minimum
information loss due to a local measurement in one of the constituents. They
generalize the one-way information deficit and contain the geometric discord as
a particular case, preserving at the same time the basic properties of the
quantum discord like reducing to the (generalized) entanglement entropy in the
case of pure states and vanishing just for classically correlated states. We
have shown that all these measures indicate the presence of long range pairwise
quantum correlations for $|B|<B_c$ in the exact ground state of these chains,
which arise essentially from the definite $S_z$ parity of such state and can be
understood in terms of the model based on the mixture of aligned states
(\ref{st}). They all reach full range at the factorizing field, where they
acquire a finite non-negligible constant value which is independent of the pair
separation or coupling range and is determined solely by the coupling
anisotropy. Such value is exactly described by the states (\ref{st}) or
(\ref{ste}), which also provide a quite reliable description of these
correlations for all $|B|<B_c$ for long range couplings, as we have seen in the
case of the fully connected model. Parity effects are then seen to be of
paramount importance for a proper description of quantum correlations in finite
quantum systems. A final comment is that the use of simple entropic forms
involving just low powers of the density matrix, like those underlying the
geometric discord $I_2$ and the cubic measure $I_3$, enables an easier
evaluation, offering at the same time an increased sensitivity of the
optimizing measurement to  changes in the type of correlation.

\section*{Acknowledgements}

The authors acknowledge support form CONICET  (N.C. and L.C. ) and CIC (R.R.)
of Argentina.


\section*{References}

\begin{thebibliography}{0}
\bibitem{NC.00}M.A.\ Nielsen and I. Chuang, {\it Quantum Computation and
               Quantum Information} (Cambridge Univ.\ Press, 2000).
 \bibitem{Ve.06}V.\ Vedral, {\it Introduction to Quantum Information Science}
 (Oxford Univ.\ Press, 2006).
 \bibitem{HR.07} S.\ Haroche and J.M.\ Raimond, {\it Exploring the Quantum}
 (Oxford Univ.\ Press, 2007).
\bibitem{QT.93}C.H.\ Bennett et al, {\it Phys.\ Rev.\ Lett.\ } {\bf 70}, 1895 (1993).
\bibitem{SD.92}C.H.\ Bennett and S.J.\ Wiesner, {\it Phys.\ Rev.\ Lett.\ } {\bf 69}, 2881 (1992).
\bibitem{JL.03}R.\ Josza and N.\ Linden, {\it Proc.\ R.\ Soc.\ } {\bf A459}, 2011 (2003);
G. Vidal, {\it Phys.\ Rev.\ Lett.\ } {\bf 91}, 147902 (2003).
\bibitem{KL.98} E.\ Knill and R.\ Laflamme, {\it Phys.\ Rev.\ Lett.\ } {\bf 81}, 5672 (1998).
\bibitem{DFC.05} A.\ Datta, S.T.\ Flammia and C.M.\ Caves,
 {\it Phys.\ Rev.\ } {\bf A72}, 042316 (2005).
\bibitem{Zu.01} H. Ollivier and W.\ H.\ Zurek,
{\it Phys.\ Rev.\ Lett.\ } {\bf 88}, 017901 (2001);
W.\ H.\ Zurek, {\it Rev.\ Mod.\ Phys.\ } {\bf 75}, 715 (2003); W.\ H.\ Zurek, Phys.\ Rev.\
{\bf A67}, 012320 (2003).
\bibitem{HV.01}L.\ Henderson and V.\ Vedral, {\it J. Phys.\ } {\bf A34}, 6899 (2001);
V.\ Vedral, {\it Phys.\ Rev.\ Lett.\ } {\bf 90}, 050401 (2003).
\bibitem{Ca.08} A.\ Datta, A.\ Shaji, and C.M.\ Caves,
{\it Phys.\ Rev.\ Lett.\ } {\bf 100}, 050502 (2008).
\bibitem{Dat.09} A.\ Datta and S.\ Gharibian, {\it Phys.\ Rev.\ } {\bf A79}, 042325
(2009).
\bibitem{SL.09}A.\ Shabani and D.A.\ Lidar, {\it Phys.\ Rev.\ Lett.\ } {\bf 102}, 100402
(2009).
\bibitem{FA.10} A.\ Ferraro et al, {\it Phys.\ Rev.\ } {\bf A81}, 052318 (2010).
\bibitem{FCOC.11}  F.F.\ Fanchini et al, {\it Phys.\  Rev.\ }  {\bf  A84}, 012313 (2011).
\bibitem{Dd.08} R.\ Dillenschneider, {\it Phys.\ Rev.\ } {\bf B 78}, 224413 (2008).
\bibitem{SA.09} M.S.\ Sarandy, {\it  Phys.\ Rev.\ } {\bf A80}, 022108 (2009).
\bibitem{MG.10} J.\ Maziero et al, {\it Phys.\ Rev.\ } {\bf A82}, 012106 (2010).
\bibitem{WR.10} T.\ Werlang and G.\ Rigolin, {\it Phys.\ Rev.\ } {\bf A81}, 044101
(2010).
\bibitem{CRC.10} L. Ciliberti,  R.\ Rossignoli, and N.\ Canosa,
{\it Phys.\ Rev.\ }  {\bf A82}, 042316 (2010).
\bibitem{Luo.08} S.\ Luo, {\it Phys.\ Rev.\ } {\bf A77}, 042303 (2008).
\bibitem{AR.10} M.\ Ali, A.R.P.\ Rau, and G. Alber, {\it Phys.\ Rev.\ } {\bf A81}, 042105 (2010);
ibid.\ {\bf 82}, 069902(E) (2010).
\bibitem{AD.11}D.\ Girolami and G.\ Adesso, {\it Phys.\ Rev.\ }  {\bf A83}, 052108 (2011).
\bibitem{Lu.08}  S.\ Luo, {\it Phys.\ Rev.\ }  {\bf A77}, 022301 (2008).
\bibitem{WPM.09}S.\ Wu, U.V.\ Poulsen and K.\ M\o lmer,
{\it Phys.\ Rev.\ }  {\bf A80}, 032319 (2009).
\bibitem{Mo.10}  K.\ Modi, T.\ Paterek, W.\ Son, V.\ Vedral and M.\ Williamson,
{\it Phys.\ Rev.\ Lett.\ } {\bf 104}, 080501 (2010).
\bibitem{DVB.10} B.\ Daki\'c, V.\ Vedral, and \v{C}.\ Brukner,
{\it Phys.\ Rev.\ Lett.\ } {\bf 105}, 160502 (2010).
\bibitem{RCC.10} R.\ Rossignoli, N.\ Canosa and L. Ciliberti,
{\it Phys.\ Rev.\ } {\bf A82}, 052382 (2010).
\bibitem{RCC.11} R.\ Rossignoli, N.\ Canosa and L. Ciliberti,
{\it Phys.\ Rev.\ } {\bf A84}, 052329 (2011).
\bibitem{Mo.11} K. Modi et al, ``Quantum discord and other measures of quantum correlations''
arXiv: 1112.6238 (2011).
\bibitem{SKB.11} A.\ Streltsov, H. Kampermann, and D.\ Bru\ss, {\it Phys.\ Rev.\ Lett.\ }
{\bf 106}, 160401 (2011).
\bibitem{MD.11} V.\ Madhok and A.\ Datta, {\it Phys.\ Rev.\ } {\bf A83}, 032323 (2011).
\bibitem{CAB.11} D.\ Cavalcanti et al, {\it  Phys.\ Rev.\ } {\bf A83}, 032324 (2011).
\bibitem{PGA.11} M.\ Piani et al, {\it Phys.\ Rev.\ Lett.\ } {\bf 106}, 220403 (2011).
 \bibitem{HH.05} M.\ Horodecki, et al,  {\it Phys.\ Rev.\ } {\bf A71}, 062307 (2005);
J.\ Oppenheim et al {\it Phys.\ Rev.\ Lett. } {\bf 89}, 180402 (2002).
\bibitem{Wh.78} H.\ Wehrl, {\it Rev.\ Mod.\ Phys.\ } {\bf 50}, 221 (1978).
\bibitem{Bha.97} R.\ Bhatia, {\it Matrix Analysis} (Springer, NY, 1997);
 A.\ Marshall and I.\ Olkin, {\it Inequalities: Theory of Majorization and its
Applications} (Academic Press, 1979).
\bibitem{CR.02} N.\ Canosa and R.\ Rossignoli,  {\it Phys.\ Rev.\ Lett.\ } {\bf  88}, 170401 (2002).
\bibitem{Kur.82} J.\ Kurmann, H.\ Thomas and G.\ M\"uller, {\it Physica }
{\bf A112}, 235 (1982).
\bibitem{Ros.04}T. Roscilde et al, {\it Phys.\ Rev.\ Lett.\ } {\bf 93}, 167203 (2004).
\bibitem{Ros.05} T.\ Roscilde et al, {\it Phys.\ Rev.\ Lett.\ }  {\bf 94}, 147208 (2005).
\bibitem{Am.06}L. Amico et al, {\it Phys.\ Rev.\ } {\bf A74}, 022322 (2006);
F. Baroni et al, {\it J.\ Phys.\ } {\bf A40} 9845 (2007).
\bibitem{RCM.08} R. Rossignoli, N. Canosa and J.M.\ Matera,
{\it Phys.\ Rev.\ } {\bf A77}, 052322 (2008).
\bibitem{GAI.09} S.M.\ Giampaolo, G.\ Adesso and F.\ Illuminati, {\it Phys.\ Rev.\ Lett.\ }
{\bf 100}, 197201
(2008); {\it Phys.\ Rev.\ } {\bf B79}, 224434 (2009).
\bibitem{RCM.09} R.\ Rossignoli, N.\ Canosa and J.M.\ Matera, {\it Phys.\ Rev.\ }
{\bf A80}, 062325 (2009);  N.\ Canosa, R.\ Rossignoli and J.M.\ Matera,  {\it Phys.\
Rev.\ } {\bf B81}, 054415 (2010).
\bibitem{GAI.10}S.M.\ Giampaolo, G.\ Adesso and F.\ Illuminatti, {\it Phys.\ Rev.\ Lett.\ }
{\bf 104}, 207202 (2010).
\bibitem{TT.11} B.\ Tomasello, D.\ Rossini, A.\ Hamma and L.\ Amico, {\it Europhys.\ Lett.\ }
{\bf 96}, 27002 (2011).
 \bibitem{CKW.00}V.\ Coffman, J.\ Kundu, and W.K.\ Wootters, {\it Phys.\ Rev.\ }
 {\bf A61}, 052306 (2000);
T.\ J.\ Osborne and F.\ Verstraete, {\it Phys.\ Rev.\ Lett.\ } {\bf 96}, 220503 (2006).
\bibitem{Wo.98} S.\ Hill and  W.K.\ Wootters, Phys.\ Rev.\ Lett.\ {\bf 78},
5022 (1997); W.K.\ Wootters, Phys.\ Rev.\ Lett.\ {\bf 80}, 2245 (1998).
\bibitem{Sch.95} B.\ Schumacher, {\it Phys.\ Rev.\ } {\bf A51}, 2738 (1995);
 C.H.\ Bennett, H.J. Bernstein, S.\ Popescu and B.\ Schumacher,
 {\it Phys.\ Rev.\ } {\bf A53}, 2046 (1996).
\bibitem{HHH.96} R.\ Horodecki and H.\ Horodecki, {\it Phys.\ Rev.\ } {\bf A54}, 1838 (1996).
\bibitem{NK.01} M.A.\ Nielsen and M.\ Kempe, {\it Phys.\ Rev.\ Lett.\ } (2001).
\bibitem{RC.02}R. Rossignoli and N. Canosa, {\it Phys.\ Rev.\ }  {\bf A66}, 042306 (2002).
\bibitem{RC.03} R. Rossignoli and N. Canosa, {\it Phys.\ Rev.\ }  {\bf A67}, 042302 (2003).
\bibitem{RW.89}  R.F.\ Werner {\it Phys.\ Rev.\ } {\bf A40}, 4277 (1989).
\bibitem{BDSW.96} C.H.\ Bennett, D.P.\ DiVincenzo, J.A.\ Smolin and
               W.K.\ Wootters, {\it Phys.\ Rev.\ } {\bf A54}, 3824 (1996).
\bibitem{Ve.02} V.\ Vedral, {\it Rev.\ Mod.\ Phys.\ } {\bf 74}, 197 (2002).
\bibitem{Ca.03}P.\ Rungta, C.M.\ Caves, {\it Phys.\ Rev.\ } {\bf A67}, 012307 (2003).
\bibitem{Ts.09}C. Tsallis, {\it J.\ Stat.\ Phys.\ } {\bf 52}, 479 (1988);
{\it Introduction to non-extensive statistical mechanics,} (Springer, NY, 2009).
\bibitem{Vi.00} G.\ Vidal, {\it J.\ Mod.\ Opt.\ } {\bf 47}, 355 (2000).
\bibitem{PF.09} A.\ De Pasquale and P.\ Facchi, {\it Phys.\ Rev.\ } {\bf A80}, 032102 (2009).
\bibitem{CR.07}N.\ Canosa and R.\ Rossignoli, {\it Phys.\ Rev.\ } {\bf A75}, 032350 (2007).
\bibitem{VPM.04} J.\ Vidal, G.\ Palacios, and R.\ Mosseri, Phys.\ Rev.\
 {\bf  A69} 022107 (2004);
S.\ Dusuel and J.\ Vidal, Phys.\ Rev.\ Lett.\ {\bf 93}, 237204 (2004).
\bibitem{DV.05}S. Dusuel and J.\ Vidal, Phys.\ Rev.\  {\bf  B71}, 224420 (2005).
\bibitem{MRC.08}J.M.\ Matera, R.\ Rossignoli and N.\ Canosa, {\it Phys.\ Rev.\ }
{\bf A78}, 012316 (2008);
{\it ibid} {\it Phys.\ Rev.\ } {\bf A82}, 052332 (2010).
\bibitem{LM.61}E.\ Lieb, T.\ Schultz and D.\ Mattis, {\it Ann.\ of Phys.\ } {\bf 16},
407 (1961).
\bibitem{ON.02}T.J.\ Osborne and M.A.\ Nielsen, {\it Phys.\ Rev.\ } {\bf A66}, 032110 (2002).
\end{thebibliography}
\end{document}